\documentstyle[12pt]{article}
\def\thefootnote{\fnsymbol{footnote}}
\newcommand{\eq}{\begin{equation}}
\newcommand{\en}{\end{equation}}
\newcommand{\eqa}{\begin{eqnarray}}
\newcommand{\ena}{\end{eqnarray}}

\newcommand{\lf}{\left}
\newcommand{\ri}{\right}

\newcommand{\sk}{\hskip .5cm }
\newcommand{\mk}{\hskip 1.cm }
\newcommand{\lk}{\hskip 1.5cm }
\newcommand{\JP}[1]{J.\ Phys.\ {\bf #1}}
\newcommand{\NP}[1]{Nucl.\ Phys.\ {\bf #1}}

\newcommand{\IJMP}[1]{Int.\ J.\ Mod.\ Phys.\ {\bf #1}}

\newcommand{\PH}[1]{Physica \ {\bf #1}}
\newcommand\bea{\begin{eqnarray}}
\newcommand\eea{\end{eqnarray}}

\begin{document}
\begin{titlepage}
\begin{flushright}
DFTT 72/98\\
ITFA 98-42\\
MS-TPI-98-22\\
September  1999
\end{flushright}
\begin{center}
{\Large\bf Universal amplitude ratios in the 2D}
\vskip 0.3cm
{\Large\bf  four state Potts model}
\end{center}
\vskip 0.8cm
\centerline{
Michele  Caselle$^a$\footnote{e--mail: caselle@to.infn.it},
Roberto Tateo$^b$\footnote{e--mail: tateo@wins.uva.nl} and
Stefano Vinti$^c$\footnote{e--mail: vinti@uni-muenster.de}
}
 \vskip 0.6cm
 \centerline{\sl  $^a$ Dipartimento di Fisica
 Teorica dell'Universit\`a di Torino and}
 \centerline{\sl Istituto Nazionale di Fisica Nucleare, Sezione di Torino}
 \centerline{\sl via P.Giuria 1, I-10125 Torino, Italy}
 \vskip .2 cm
 \centerline{\sl $^b$  Universiteit van Amsterdam, Instituut voor Theoretische
   Fysica,  }
 \centerline{\sl  1018 XE Amsterdam, The Netherlands }
 \vskip .2 cm
 \centerline{\sl $^c$ 
  Institut f\"ur Theoretische Physik I, Universit\"at  M\"unster,}
 \centerline{\sl Wilhelm-Klemm-Str. 9, D-48149 M\"unster, Germany}
 \vskip 0.6cm

\begin{abstract}
We present a  Monte Carlo
 study of various universal amplitude
ratios of  the two dimensional $q=4$ Potts model. We simulated
the model close to criticality in both phases
taking care to keep the 
systematic errors, due to finite size effects
and logarithmic corrections in the  scaling functions, under control.
Our results are 
compatible 
 with those  recently obtained using the form-factors  approach and 
with the existing low temperature series for the model.
\vskip0.1cm
\end{abstract}

\begin{flushleft}
{\sl PACS: 75.10.H , 11.10.J }\\
{\sl Keywords: 4-state Potts, Monte Carlo, universal  amplitude ratios.}
\end{flushleft}
\end{titlepage}

\setcounter{footnote}{0}
\def\thefootnote{\arabic{footnote}}

\section{Introduction}
One of the  peculiar features of
the two dimensional four state Potts model
 is the presence of a marginal field which
leads to universal multiplicative logarithmic corrections to the scaling laws. 

These corrections can be exactly evaluated~\cite{ns,cns}  but, as  often
happens when dealing with marginal fields, they are accompanied 
by still large subleading non-universal contributions which  completely  
mask the
 scaling  behaviour  of the system (at least for those values of 
the correlation length which can be
 reached in standard simulations).

This is  different from the
 behaviour,  for instance, of the 3d Ising model where
even at moderate 
values of the correlation length,
the  non-universal 
corrections~\cite{ch}  
give  very small contributions; thus  they can  be safely taken 
into account  by adding
to the scaling functions only the first 
non-universal term.

In the present case, instead, these corrections
are so large that 
the reliability of the  
fits, in which only the first subleading 
term is taken into account, becomes   questionable.
On the other hand, with  current numerical precision, it is almost
impossible to add further corrections 
without losing any predictive power in the fits. 

This makes the numerical study of the four state Potts model one of the most 
difficult tasks in the context of Monte Carlo simulations of  two dimensional
spin models. 

This  problem 
was
recently  addressed by  J.Salas and A.D.Sokal in~\cite{ss2}
by  
extending the RG  analysis of~\cite{ns,cns} up to  third order in the fields.
They 
succeeded in obtaining
the universal leading corrections to the scaling, 
 which turn out to be 
additive terms of the generic form $log~log/log$. 
The improved scaling functions were  then tested  by looking
at the critical finite size properties of the model, and 
an  improvement of the  scaling behaviour of the   data
was observed which 
however turned out  to
be still  affected by large non-universal $1/log$ terms.

In this paper we return to  this problem by looking at various universal
amplitude combinations of the model~\cite{ahp}. We find 
an improvement in the scaling behaviour of our data if the universal 
contributions evaluated in ~\cite{ss2} are taken into account.

However, as in~\cite{ss2}, this is not enough to describe the data; 
non-universal corrections  must be considered     and 
final results crucially  depend on the type of  terms included    in the   
scaling functions. 
We consider this one of the most delicate aspects of this paper. 
For this reason we   described, as precisely as possible, 
the procedure  used to construct the  scaling functions (see sect. 7) 
and   included in the paper, besides  final estimates for the  
amplitude  ratios, also the direct results of the Monte Carlo simulation 
(see tables 1,2 and 3) so that the reader can use the data to study 
alternative scaling functions and possibly find a clever
way to control the systematic errors involved in the truncation that 
we suggest.

Fortunately for the present problem we have an independent way to 
test our results. In fact, thanks to a recent work by J.L.Cardy and 
G.Delfino~\cite{cd}, precise estimates for  amplitude combinations 
are now available in the continuum  limit. 
The  relatively good agreement  found between our estimates
and those of~\cite{cd} make us confident of the reliability of our 
results and at the same time strongly supports the correctness of the 
form-factors  derivation of ref.~\cite{cd}.

This paper is organized as follows. In sect.~2 we give a general 
introduction to
the $q$-state Potts model and we summarize a few known facts concerning 
the phase diagram  and its scaling limit 
description in the framework of conformal and 
perturbed conformal field theory.
In sects. 3 and 4  we introduce and discuss the observables and the 
amplitude ratios in which we are interested. 
In sect.~5 and 6 we describe the Monte Carlo simulation and test the 
results by comparing them with existing low temperature series and by 
imposing the duality relations on the internal energy and the specific heat. 
Sect.~7 is devoted to the study of the scaling behaviour of the observables 
and to  extract the best estimates for the amplitude combinations. Sect.~8 
is devoted to a comparison of the results with those of ref.~\cite{cd} while 
in sect.~9 we have  collected some concluding remarks.

\section{The Model}
 
We study the four state Potts model in two dimensions on a simple square 
lattice. 
The action is given by
\eq
 S_{Potts} = - \beta \sum_{<x,y>} \delta_{s(x), s(y)} \sk ,
\label{1.1}
\en
where
 the field variable $s(x)$ takes the values $0,1,2,3$;
 $~~x\equiv(x_0,x_1)$ labels the sites of the lattice and the notation 
$<x,y>$  indicates that the sum is taken on  nearest neighbor sites 
only. The $\delta$ function is defined as usual: $\delta_{a,b}=1$ if $a=b$ and
$0$ otherwise. 
 The coupling $\beta$ is related to the temperature in the 
standard  way
 $\beta\equiv \frac{1}{kT}$. In the following
we shall always consider lattices of 
equal extension $L$ and periodic boundary conditions in both directions.

Several results are known exactly for this model.
The action eq.~(\ref{1.1}),  is invariant under the permutation 
group $S_4$. However in the low temperature phase this symmetry is 
spontaneously
broken to $S_3$. The two phases are 
related by duality
and separated by a second order phase 
transition located at  $\beta_c\equiv \frac{1}{kT_c}=\log(3)=1.098612$.
The dual coupling
$\tilde\beta$ is related to the original coupling $\beta$ by
\eq
\tilde\beta=-\log\left(\frac{1-e^{-\beta}}{1+3e^{-\beta}}\right) \sk,
\en
and the fixed point of this relation is  the critical coupling $\beta_c$.

It is useful to introduce the variables
\eq
\sigma_\alpha(x)=\delta_{s(x),\alpha}-\frac14 \mk , \mk  \alpha=0,1,2,3
\sk .
\en
It is easy to see that $<\sigma_\alpha>=0$  $\forall\alpha$ in the high
temperature phase and that they all become different from zero 
in the low temperature phase. In particular for one of the four values of the
spin   (which we shall call in the following ``majority spin'')
$<\sigma_\alpha>~>~0$, while for the three other values we have 
 $<\sigma_\alpha>~<~0$.

It has been  shown (cf.  \cite{fk})   that the partition
function $Z(T,q)$ for the $q$-state Potts  models ($q$ integer)
 on a square lattice $\Lambda$  can be written as  
\eq
Z(T,q) = \sum_{\{G\}} (e^{\beta}-1)^N q^{\nu} \sk .
\label{par}
\en
The sum in eq.~(\ref{par}) runs  over all the graphs $G$
on $\Lambda$,  $C$ is the number of connected components
(including isolated sites) in $G$, and $N$ is the number of bounds
on the lattice edges.
For a better   understanding of    some of the peculiar features
 of the four-state Potts 
model it is  convenient to consider the phase diagram of the whole 
family of models,  defined for arbitrary   $q>0$.
Eq.~(\ref{par})  provides an expression suitable for  
extending  the definition of $Z(T,q)$ to non integer values of $q$.
The $q$-state   Potts model 
undergoes a phase transition at 
\eq
\beta_c=\log(\sqrt{q}+1) \sk .
\en
Below this temperature the system is in its $S_q$-broken symmetry phase
whereas above it the system is  fully disordered.   
The  transition  at $T=T_c$  is  first order for $q>4$ but
becomes second order for $q \leq 4$,
in the latter case the model gets renormalized  into a conformal field theory 
with 
central charge~\cite{df} 
\eq
c=1- { 6 \over (l-1) l} \sk ,
\en
where $l$ is related to $q$ by 
\eq
{  2 \pi  \over l}= \arctan \lf(  { \sqrt{4 q- q^2} \over (q-2)} \ri) \sk.
\label{tq}
\en 
In the scaling  limit, at   rational points  $l=p/(p'-p)$,
the  thermal field $\varepsilon$ rescales   
with scaling dimension~\cite{nie}
\eq
\Delta_{\varepsilon}={1 \over 2} \lf( 1+ {3 \over {l-1}} \ri) \sk ,
\en
hence it  can be  identified with 
the  operator  $\phi_{21}$  in the $M_{p,p'}$ minimal conformal 
model.
Notice also that eq.~(\ref{tq})
shows that two square-root branch points at $q=0$ 
and $q=4$ are present.
At $q=0$ the thermal operator is marginal, in the analytically-continued 
second branch 
it  becomes
irrelevant and  the critical point has  moved
into
the antiferromagnetic region.  The physics in this sector
is certainly very interesting 
but it is slightly less   relevant for our 
current interests. More related to  the subject of this paper
 is, instead, the  physical meaning of the second  
branch point at  $q=4$. Let us consider  a further variant of the
model  in which  
vacancies are  allowed, and correspondingly a chemical potential $\mu$ is 
introduced. In the sector $0<q<4$ with  $\mu$ negative or
sufficiently small, the additional dilution 
field turns out to be 
irrelevant and  the system  still  undergoes a second order phase 
transition
in  the same universal class of the pure Potts model.
Near the transition point  the dilution field scales  with a
conformal dimension
\eq
\Delta_{\mu}=2+ {4 \over {l-1} } \sk ,
\en
and it can be  identified  with   
primary conformal operator $\phi_{31}$. 

At $q=4$, the  dilution  field $\phi_{\mu}$ becomes marginal;
along the critical RG flow its  slow  rate of
 disappearance cause now multiplicative logarithm corrections 
to the critical behaviour.  
{}From eq.~(\ref{tq}) we also see  that the net result  
of the entrance  in the second branch consists of a negation of
 $l$. Hence  at the  same value of $q$, but on the second branch
we now have a conformal field theory (CFT) with central charge 
\eq
c=1- {6 \over l(l+1)} \sk ,
\en
with thermal   and  dilution fields with  dimensions
\eq
\Delta_{\varepsilon}={1 \over 2} \lf( 1- {3 \over {l+1}} \ri)
\sk , \sk
\Delta_{\mu}=2- {4 \over {l+1}} \sk.
\en
These two fields can  now be    respectively  identified   with 
the relevant conformal operators   $\phi_{12}$ and $\phi_{13}$.
In conclusion, at fixed  $q < 4$, the  phase diagram in the plane 
$(\mu,T)$ is  as follows: 
if $\mu$ is negative or sufficiently small  then at $T=T_{c}(\mu)$
the system 
undergoes a second order phase transition in the universal class of 
the pure Potts
model, whereas  when $\mu$ is large the transition 
becomes of the first order. On  the critical line $(\mu,T_c(\mu))$
the point marking the change of critical behavior is the tricritical
point.
{}From the scaling  quantum field theory point of view, the picture looks also
consistent with the one described above.
First   notice that  perturbation of the conformal field theory 
 with either the thermal or the 
dilution field  is 
integrable~\cite{zam0} and that  
the associated quantum  field theories have been 
the subject of very deep studies~(see for example 
\cite{sm,sm1,bl,zam1,zam2,zam3}). 
The thermal operator $\varepsilon$  drives the system into a massive 
ordered or disordered  phase depending on the sign of the perturbing 
parameter~\cite{zam1,zam2,zam3}.
The operator $\phi_\mu$  instead moves  the tricritical model either into
a massive phase 
(a line of first-order transitions)   
or into a critical massless phase.
The IR fixed point of the latter is the Potts-model CFT and the 
two less irrelevant attracting operators are the fields 
$\phi_{31}$ and $T\bar{T}$~\cite{zam1,zam2,zam3}.

Quantum  reductions of  the Izergin-Korepin model at rational points, giving 
the S-matrix elements for the $\phi_{12}/\phi_{21}$ perturbations, were first
obtained by F.A.Smirnov~\cite{sm}. Subsequently using a somehow different kink
basis,
L.Chim and A.B.Zamolodchikov~\cite{cz} defined alternative scattering elements 
for the model. 
The latter formulation, being more suitable for analytic continuation at
arbitrary values of $q$,  has been used by J.L.Cardy and G.Delfino~\cite{cd} 
to make predictions about the values of some universal amplitudes.
The method used is a variant  of the form-factors approach to the
correlation functions proposed in~\cite{kw}.

\section{The observables}

\subsection{Magnetization}
The magnetization of a given configuration is  defined as:
\begin{equation}
m = \frac{1}{V} \sum_x \sigma_{\alpha_m}(x) \sk ,
\end{equation}
where $V\equiv L^2$ is the volume of the lattice and $\alpha_m$ is the value
of spin 
corresponding to the majority of the spins. 
However, in a finite volume at arbitrary finite low   temperature 
the $S_4$ symmetry of the model is not  spontaneously broken.
Practically this means that, in the
simulation sample, configurations with
all the four possible values of $\alpha_m$ appear with equal probability. 
In order to obtain a low temperature  non-vanishing  magnetization 
a  magnetic field $h$
that explicitly  breaks    the symmetry,  must be  coupled to the system.
The thermodynamic limit at non zero  $h$ 
 should be taken first, then the limit
of vanishing magnetic field could be  performed. 
However it is difficult to follow this route in a numerical study. 
An alternative, commonly adopted,
approach  is to identify $\alpha_m$
in each configuration by counting the spins belonging to the four
possible  values of
$\alpha$ and then extracting the majority one. 
This procedure works in a satisfactory way if
the lattice size and coupling constants are such that the probability of finding
interfaces among different
vacua is negligible. We carefully chose our lattice sizes so as to satisfy
this bound
\footnote{Let us note, as a side remark,
 that this procedure, in the Ising
 case is equivalent to the choice
\[
<m>~\equiv ~\lim_{L\to\infty} <|m|> \lk (m = \frac{1}{V} \sum_i s_i ) \sk,
\label{m1}
\]
where the $s_i$'s are in this case Ising spins.

The finite size behaviour of this observable was carefully 
studied in~\cite{tb}. It was shown that  this  choice converges to the
infinite volume value better
than any other existing proposal
 and that the asymptotic, infinite volume,
value is reached for lattices of size $L>\sim 8\xi$, where $\xi$ denotes the
correlation length. In our simulations we always used lattice
sizes much larger than this threshold.}.

In the following we shall assume for simplicity that 
$\alpha_m=0$ is the value of the majority spin and shall denote the remaining
three values with roman indices $i,j,k,...=1,2,3$.

Close to criticality  and at $
t\equiv\frac{\beta_c-\beta}{\beta_c}
<0$, the magnetization scales 
as~\cite{ss2}

\begin{equation}
\label{defb}
 <m> \sim \; B \; (-t)^{\frac{1}{12}}   
 (- \log(-t) )^{-\frac{1}{8}}  \,
  \left[ 1 - \frac{3}{16} 
         \frac{\log(-\log (-t))}{ -\log (-t)}
         + O\!\left( \frac{1}{\log (-t)} \right) \right]~. 
\end{equation}

\subsection{Magnetic susceptibility}
The susceptibility 
\begin{equation}
\chi = \frac{\partial <m> }{\partial H}
\end{equation}
gives the response of the magnetization 
to an external magnetic
field and it can be  
expressed in 
terms of moments of the magnetization  
\begin{equation}
\chi = V \left( <m^2> - <m>^2 \right) \sk.
\end{equation}
In the high temperature phase this means
\begin{equation}
\chi = V <m^2> ~~~=V <\sigma_\alpha^2> \sk,
\end{equation}
where $\alpha$ is any one of the four values $(0,1,2,3)$\footnote{Notice however
that, inspired by the analogy with the Ising model or by the embedding 
in the AT
model, different definitions of the order parameter (hence of the mean
magnetization in the symmetric phase) are possible. For instance
\[
<m_1>= <\sigma_\alpha-\sigma_\beta>
\]
with $\alpha\neq\beta$
or
\[
<m_2>= <\sigma_\alpha-\sigma_\beta+\sigma_\gamma-\sigma_\delta>
\]
with $\alpha\neq\beta\neq\gamma\neq\delta$.
The corresponding susceptibilities are related to $\chi$ by simple
multiplicative constants. In comparing our results with those 
of~\cite{ss1,ss2} one must take into account this different normalization.}.

In the  broken symmetric  phase,
depending on the choice of coupling the external magnetic field to the
majority spin or to one (or more) of the other values,
two  kinds of susceptibilities can be defined~\cite{fs}
\bea
\chi_l&=& <\sigma_0^2> \lk \lk  \mk\hbox{longitudinal}~~, \\
\chi_t&=& <(\sigma_i-\sigma_j)^2> \sk (i\neq j) \mk \hbox{transverse} \sk.
\eea
In this paper we    concentrate on the  longitudinal susceptibility  $\chi_l$.

Close to the critical temperature $\chi$ and $\chi_l$ 
scale as\footnote{Notice that there
is a misprint in the analogue of this equation in ref.~\cite{ss2}.}
\begin{eqnarray}
\label{defc}
\chi \sim \Gamma_+
(t)^{-\frac{7}{6}} (-\log|t| )^{\frac{3}{4}}   
  \left[ 1 + \frac{9}{8}
         \frac{\log(-\log |t|)}{-\log |t|}
         + O\!\left( \frac{1}{\log |t|}\right) \right] &(t>0)&~,  \nonumber \\
\chi_{l} \sim \Gamma_{-}
(-t)^{-\frac{7}{6}} (-\log|t| )^{\frac{3}{4}}   
  \left[ 1 + \frac{9}{8}
         \frac{\log(-\log |t|)}{-\log |t|}
         + O\!\left( \frac{1}{ \log |t|}\right) \right] \;  &(t<0)&~.  
\nonumber \\
\end{eqnarray}

\subsection{Internal energy and specific heat}
The internal energy is defined as
\eq
E=\frac{1}{2V} \sum_{<x,y>} \delta_{s(x), s(y)} \sk, 
\en
the specific heat 
\begin{equation}
C\equiv\frac{d<E>}{d\beta} = 2V \left( <E^2> - <E>^2 \right) \sk.
\end{equation}
Duality relates both  internal energy and specific heat 
in the low temperature phase to those in  the high temperature phase.
The relations are
\eq
(1-e^{-\beta}) E(\beta)=1 - 
(1-e^{-\tilde\beta}) E(\tilde\beta) \sk,
\label{edual}
\en
\eq
C(\beta)(1-e^{-\beta})^2+
e^{-\beta}(1-e^{-\beta}) E(\beta)= 
C(\tilde\beta)(1-e^{-\tilde\beta})^2+
e^{-\tilde\beta}(1-e^{-\tilde\beta}) E(\tilde\beta).
\label{cdual}
\en
Close to the critical temperature we have
\begin{eqnarray}
\label{defe}
C \sim A_+ \
 (t)^{-\frac{2}{3}} (- \log|t| )^{-1}  
    \left[ 1 - \frac{3}{2}
         \frac{\log(-\log |t|)}{ -\log |t|}
         + O\!\left( \frac{1}{ \log |t|}\right) \right] 
   &(t > 0)&~, \nonumber \\
C \sim A_- \
 (-t)^{-\frac{2}{3}} (- \log|t| )^{-1}  
    \left[ 1 - \frac{3}{2}
         \frac{\log(-\log |t|)}{ -\log |t|}
         + O\!\left( \frac{1}{ \log |t|}\right) \right] 
  &(t < 0)&~. 
\nonumber \\
\end{eqnarray}

\subsection{Second moment correlation length}
We consider the decay of so-called time-slice
correlation functions. The magnetization of a time-slice is given by
\begin{equation}
 S_{\alpha}(x_0) = \frac{1}{L}\sum_{x_1} \sigma_{\alpha}(x_0,x_1) \sk.
\label{timeslice}
\end{equation}
Let us define the correlation function 
\begin{equation}
G_{\alpha\beta}(\tau) = \sum_{x_0} \left\{ \langle
 S_{\alpha}(x_0) S_{\beta}({x_0}+\tau) \rangle - 
\langle S_\alpha(x_0) \rangle~\langle S_\beta(x_0) \rangle \right\} \sk .
\label{gg}
\end{equation}
For any choice of the indices $\alpha$ and $\beta$ in~(\ref{gg}), 
the dominant large distance behaviour 
of $G(\tau)$  is  
dominated by the lowest mass of the model:
\begin{equation}
G(\tau) \propto \exp(-\tau/\xi) \sk ,   
\end{equation}
where $\xi$ is the exponential correlation length and coincides  with 
the inverse of the lowest 
 mass of the model. 
However, at low temperature,  the rich structure of the 
spectrum   (cf.~\cite{cd}) can mask this asymptotic behaviour. 
For this reason, in the following,  we shall concentrate  on $G_{00}$ 
which has the ``neatest''  asymptotic behaviour. 
This will also allow us to directly
compare our results with those of \cite{cd}.  
In the high temperature  phase the lowest mass is well separated 
from all other excitations in the spectrum and extracting the
exponential  correlation length is much simpler.

Close to criticality the behaviour of the correlation length is governed 
by the scaling laws
 \begin{eqnarray}
\label{deff}
\xi \sim f_+\, (t)^{-\frac{2}{3}} \, (- \log|t|)^{\frac{1}{2}} \,  
        \left[ 1 + \frac{3}{4}
               \frac{\log(-\log |t|)}{ -\log |t|} +
               O\!\left( \frac{1}{\log |t|}\right) \right] 
  &(t>0)&~,  \nonumber \\
\xi \sim f_-\, (-t)^{-\frac{2}{3}} \, (- \log|t|)^{\frac{1}{2}} \,  
        \left[ 1 + \frac{3}{4}
               \frac{\log(-\log |t|)}{ -\log |t|} +
               O\!\left( \frac{1}{\log |t|}\right) \right] 
  &(t<0)&~.  \nonumber \\
 \end{eqnarray}
We are also  interested in the 
 second moment correlation length $\xi_2$, to evaluate $\xi_2$
 we used the estimator~\cite{ss1,ss2}
\eq
\xi^{(2)} \;\equiv\; { \left( {\chi \over F}  -1 \right)^{\frac{1}{2}} \over 
                   2\sin (\pi/L) 
                 } \sk ,  
\label{xi2}
\en
where $\chi$ is the susceptibility 
and $F$ is the Fourier transform of the correlation function
 at the smallest nonzero momentum $(2\pi/L, 0)$.

Notice that the susceptibility can be rewritten as  
zero momentum Fourier transform of the correlation function, hence, 
in order to have a consistent definition, the same correlation function 
must be
chosen in both $\chi$ and $F$.
In the  low temperature regime we
are interested in setting the longitudinal 
susceptibility
\eq
\chi_l  \equiv \sum_x    G_{0,0}(x)
\label{u1}
\en
in eq.~(\ref{xi2}), hence, we  shall study
\eq
F \equiv \sum_x  e^{2i\pi x_1/L}  G_{0,0}(x) \sk .
\label{u2}
\en
$\xi_2$ is a very popular approximation for the exponential correlation length
since, 
in Monte Carlo simulations,
its numerical evaluation  is much simpler
than that of $\xi$.

Moreover it is the length scale
which is directly observed
in scattering experiments. 
It is important
to stress that 
 $\xi_{2}$  and $\xi$ are not fully equivalent~(cf. \cite{ch} ),
even though their 
 critical behaviours are 
the same up to a multiplicative factor.
In particular,  
the ratio $\xi/\xi_{2}$ 
gives an idea of the density of the lowest states of
the spectrum. If the lower excited states  are well separated the ratio 
is  almost
one, whereas a
significantly bigger ratio indicate a denser distribution of
states.  Furthermore, different choices of the correlation function
in eq.s~(\ref{u1},\ref{u2})  lead to different values  of $\xi_{2}$, while
the exponential correlation length is
always the same. 
A careful study of these differences can give several
information on the spectrum of the theory.  
We shall call the
critical amplitudes $f_{2,\pm}$  for $\xi_2$ to distinguish them from $f_\pm$.
\section{Amplitude ratios}
We are interested in the following amplitude ratios
\eq
R_\chi=\frac{\Gamma_+}{\Gamma_{-}} \mk, \mk
R_{\xi,2}=\frac{f_{2,+}}{f_{2,-}} \sk ,
\label{r1}
\en
and the following amplitude combinations
\eq
R_1=\frac{\Gamma_-}{f_{2,-}^2B^2} \mk , \mk
R_2=\frac{\Gamma_+}{f_{2,+}^2B^2} \sk , 
\label{r2}
\en
which are scale invariant thanks to the  
(hyper)scaling   relations among the
critical exponents
\eq
\alpha+2\beta+\gamma=2 \sk, \lk
d\nu=2-\alpha \sk.
\en
We are also interested in the combinations
\bea
R^+_c=\frac{A_+\Gamma_+}{B^2} \sk &,& \sk
R_\xi^+=\sqrt{A_+}f_{2,+} \mk , \label{r3} \\
R^-_c=\frac{A_-\Gamma_-}{B^2} \sk &,& \sk
R_\xi^-=\sqrt{A_-}f_{2,-} \mk,
\label{r4}
\eea
which have  particularly interesting behaviours in the 4 state Potts model (see
below).
We shall  neglect the amplitude ratio  $A_+/A_-$ which is trivially 1 due
to duality~\footnote{In principle this result could 
be used to test our simulation. But
such test is completely equivalent to the test of the duality relation 
eq.~(\ref{cdual}) that we perform in tables~4 and 5.}.

\section{The simulations}

We produced a standard cluster algorithm using both the Wolff single cluster 
update and the Swendson Wang cluster update. 
After preliminary tests, we used the latter algorithm for our high statistic 
simulations. 

To check our program we made comparisons of the MC results with the exact 
solution on a $3^2$ lattice and with the Salas and Sokal 
(ref.s~\cite{ss1,ss2}) 
results 
at the critical point on a $16^2$ lattice, with a comparable statistics.

We simulated the 4 state Potts model in the high and low temperature phases for
16 values of the couplings which were chosen  exactly as dual pairs. 
This choice allowed us first, to
perform a very stringent test on our estimates for the thermal observables 
which must be related by duality and second, to obtain a direct estimate of
some amplitude ratios. The results for the observables in which we are
interested are reported in tables~1,2 and 3.

Lattice sizes were chosen large enough to make finite size effects negligible
within our statistical errors.
After a preliminary test on the  finite size  behaviour of 
our observables we chose  $L > 10~\xi$ in the high temperature phase 
and $L > 20~\xi$ 
in the  low temperature phase, a default size $L=120$ was taken for small 
values of $\xi$.
In each simulation the number of measurements 
was $2\cdot 10^7$. Each measurement was
separated from the next one by two Swendsen-Wang updates.
A standard jacknife procedure has been used to analyze statistical errors.
%
%%%%%%%%%%%%%%%%%%%%%%%%%%%%%%%%%%%%%%%%%%%%%%%%%%%%%%%%%%%%%%%
\begin{table}[htb]
\begin{center}
\begin{tabular}{|c|c|c|c|c|c|}
\hline
$\beta$ &$L$  & $\xi_{2nd}$ & $E$& $C$  & $\chi$    \\
\hline
1.06722 & 120 & 6.44(5)& 0.64604(6)  & 1.571(12) & 11.83(2)  \\
1.07722 & 120 & 8.46(4)& 0.66318(6)  & 1.892(13) & 18.78(3)  \\
1.07972 & 120 & 9.25(3)& 0.66808(6)  & 2.018(14) & 21.80(4)  \\
1.08222 & 120 & 10.26(3)& 0.67337(6) & 2.189(15) & 25.90(5)  \\
1.08472 & 120 & 11.60(3)& 0.67909(6) & 2.405(15) & 31.69(6)  \\
1.08722 & 180 & 13.37(5)& 0.68544(6) & 2.614(22) & 40.31(9)  \\
1.08972 & 180 & 15.94(5)& 0.69232(6) & 3.033(24) & 54.05(13) \\
1.09222 & 240 & 20.33(7)& 0.70056(6) & 3.574(35) & 80.88(24) \\
\hline
\end{tabular}
\caption{\sl Results in the high temperature
phase.}
\label{spinht}
\end{center}
\end{table}
%%%%%%%%%%%%%%%%%%%%%%%%%%%%%%%%%%%%%%%%%%%%%%%%%%%%%%%%%%%
%
%%%%%%%%%%%%%%%%%%%%%%%%%%%%%%%%%%%%%%%%%%%%%%%%%%%%%%%%%%%
\begin{table}[htb]
\begin{center}
\begin{tabular}{|c|c|c|c|c|}
\hline
$\beta$ &$L$ &  $\xi_{2nd}$ & $E$& $C$      \\
\hline
1.130500 & 120  & 2.85(5)& 0.850888(5)  & 1.3718(10)  \\
1.120231 & 120  & 3.69(5)& 0.835061(7)  & 1.7474(16)  \\
1.117684 & 120  & 4.00(4)& 0.830453(8)  & 1.8851(17)  \\
1.115135 & 120  & 4.45(4)& 0.825436(8)  & 2.0516(20)  \\
1.112597 & 120  & 5.01(4)& 0.820001(9)  & 2.2604(24)  \\
1.110065 & 120  & 5.77(4)& 0.813951(10) & 2.5274(28)  \\
1.107540 & 120  & 6.98(4)& 0.807100(12) & 2.9048(40)  \\
1.105020 & 180  & 8.89(7)& 0.799095(9)  & 3.4872(49)  \\
\hline
\end{tabular}
\caption{\sl Results in the low temperature
phase: thermal observables and correlation lengths.}
\label{spinlt1}
\end{center}
\end{table}
%%%%%%%%%%%%%%%%%%%%%%%%%%%%%%%%%%%%%%%%%%%%%%%%%%%%%%%
%
%
%%%%%%%%%%%%%%%%%%%%%%%%%%%%%%%%%%%%%%%%%%%%%%%%%%%%%%%
\begin{table}[htb]
\begin{center}
\begin{tabular}{|c|c|c|c|}
\hline
$\beta$ &$L$ & $m$ & $\chi_l$        \\
\hline
1.130500 & 120 &  0.633288(7)  & 1.2180(9)   \\
1.120231 & 120 &  0.612874(11) & 2.1072(23)  \\
1.117684 & 120 &  0.606331(13) & 2.5080(26)  \\
1.115135 & 120 &  0.598834(14) & 3.0572(36)  \\
1.112597 & 120 &  0.590219(16) & 3.8447(51)  \\
1.110065 & 120 &  0.579950(19) & 5.0315(72)  \\
1.107540 & 120 &  0.567272(26) & 7.053(14)   \\
1.105020 & 180 &  0.550686(23) & 10.987(19)  \\
\hline
\end{tabular}
\caption{\sl Results in the low temperature
phase: magnetic observables.}
\label{spinlt2}
\end{center}
\end{table}
%%%%%%%%%%%%%%%%%%%%%%%%%%%%%%%%%%%%%%%%%%%%%%%%%%%%%
%
\section{Analysis of the results}
\subsection{Energy and specific heat}
By using eq.~(\ref{edual}) we have a non trivial test of our estimates both for
the energy and for the specific heat. In table~4 we compare the results for 
the
internal energy in the high temperature phase with those obtained using 
eq.~(\ref{edual}) and the values measured with the dual coupling, 
at low temperature as input. A similar comparison for the specific heat 
can be found in
table~5.
%
%%%%%%%%%%%%%%%%%%%%%%%%%%%%%%%%%%%%%%%%%%%%%%%%%%%%%%
\begin{table}[htb]
\begin{center}
\begin{tabular}{|c|c|c|}
\hline
$\beta$  & HT & LT+eq.(\ref{edual})      \\
\hline
1.06722 & 0.64604(6)  &  0.646061(5)   \\
1.07722 & 0.66318(6)  &  0.663179(7)   \\
1.07972 & 0.66808(6)  &  0.668074(8)   \\
1.08222 & 0.67337(6)  &  0.673365(8)   \\
1.08472 & 0.67909(6)  &  0.679055(9)   \\
1.08722 & 0.68544(6)  &  0.685341(10)  \\
1.08972 & 0.69232(6)  &  0.69240(1)    \\
1.09222 & 0.70056(6)  &  0.70060(1)    \\
\hline
\end{tabular}
\caption{\sl Comparison between the internal energy measured in the simulation
at high temperature (second column) and the values obtained,
 from the internal
energy measured at low temperature, using  the duality 
relation~{\protect (\ref{edual})} (third column).}
\label{comp1}
\end{center}
\end{table}
%%%%%%%%%%%%%%%%%%%%%%%%%%%%%%%%%%%%%%%%%%%%%%%%%%%%%%%
%
%%%%%%%%%%%%%%%%%%%%%%%%%%%%%%%%%%%%%%%%%%%%%%%%%%%%%%%
\begin{table}[htb]
\begin{center}
\begin{tabular}{|c|c|c|}
\hline
$\beta$  & HT & LT+eq.(\ref{cdual})      \\
\hline
1.06722  & 1.571(12) & 1.555(1) \\
1.07722  & 1.892(13) & 1.903(2) \\
1.07972  & 2.018(14) & 2.032(2) \\
1.08222  & 2.189(15) & 2.191(2) \\
1.08472  & 2.405(15)&  2.391(3) \\
1.08722  & 2.614(22) & 2.646(3) \\
1.08972  & 3.033(24) & 3.011(4) \\
1.09222  & 3.574(35) & 3.579(5) \\
\hline
\end{tabular}
\caption{\sl Comparison between the specific heat  measured in the simulation
at high temperature (second column) and the values obtained, from the internal
energy and the specific heat 
measured at low temperature, using the duality 
relations~{\protect (\ref{edual},\ref{cdual})} (third column).}
\label{comp2}
\end{center}
\end{table}
%%%%%%%%%%%%%%%%%%%%%%%%%%%%%%%%%%%%%%%%%%%%%%%%%%%
%
\subsection{Magnetization and Susceptibility}
In tables~6 and 7  we compare our results for the magnetization and the
low temperature longitudinal susceptibility
 with the series of ref.~\cite{beg}. 
Both for the magnetization and for
the susceptibility we used the diagonal Pade' approximant.
%
%%%%%%%%%%%%%%%%%%%%%%%%%%%%%%%%%%%%%%%%%%%%%%%%%%%
\begin{table}[htb]
\begin{center}
\begin{tabular}{|c|c|c|}
\hline
$\beta$ &our MC & series       \\
\hline
1.130500 &  0.633288(7)  & 0.633275  \\
1.120231 &  0.612874(11) & 0.612863  \\
1.117684 &  0.606331(13) & 0.606310  \\
1.115135 &  0.598834(14) & 0.598853  \\
1.112597 &  0.590219(16) & 0.590268  \\
1.110065 &  0.579950(19) & 0.580154  \\
1.107540 &  0.567272(26) & 0.567899  \\
1.105020 &  0.550686(23) & 0.552456  \\
\hline
\end{tabular}
\caption{\sl Comparison of our Monte Carlo results for the magnetization with 
a Pade' resummation of the series of ref.~{\protect \cite{beg} }.}
\label{spinlt3}
\end{center}
\end{table}
%%%%%%%%%%%%%%%%%%%%%%%%%%%%%%%%%%%%%%%%%%%%%%%%%%%%%
%
As expected, the agreement, which is rather good far from the critical point
becomes worse and worse as $\beta_c$ is approached. Notice however that we used
the simplest possible resummation technique, more sophisticated approaches like
the double biased IDA of ref.~\cite{lf} could 
give better results and could also
give a way to estimate the systematic errors involved in the truncation and
resummation of the series (for an attempt in this direction in the case of the
3d Ising model  see for instance~\cite{ch})\footnote{In comparing our values of
the magnetization with those of ref.~\cite{beg} one must notice that there is a
factor 4/3 between the two definition of magnetization. On the contrary there is
complete agreement between the two
definitions for the longitudinal susceptibility.}.
%
%%%%%%%%%%%%%%%%%%%%%%%%%%%%%%%%%%%%%%%%%%%%%%%%%%%%%%
\begin{table}[htb]
\begin{center}
\begin{tabular}{|c|c|c|}
\hline
$\beta$ &our MC & series        \\
\hline
1.130500  &  1.2180(9)  & 1.2187  \\
1.120231  &  2.1072(23) & 2.1241  \\
1.117684  &  2.5080(26) & 2.5395  \\
1.115135  &  3.0572(36) & 3.1179  \\
1.112597  &  3.8447(51) & 3.9661  \\
1.110065  &  5.0315(72) & 5.3242  \\
1.107540  &  7.053(14)  &  7.795  \\
1.105020  & 10.987(19)  & 13.647  \\
\hline
\end{tabular}
\caption{\sl Comparison of our Monte Carlo results for $\chi_l$ with 
a Pade' resummation of the series of ref.~{ \protect \cite{beg} }.}
\label{spinlt4}
\end{center}
\end{table}
%%%%%%%%%%%%%%%%%%%%%%%%%%%%%%%%%%%%%%%%%%%%%%%%%%%%%%
%
In the case of the susceptibility the discrepancy between the results from the 
series expansion and our Monte Carlo are larger and only the first 
value of beta
agree within the errors. It is clear however that we are pushing the series to
their limit of validity and in fact, looking at nondiagonal Pade'
approximants, one sees very large fluctuations (much larger than in the case of
magnetization) in the series estimates.

\section{Scaling behaviour}

Let us now address the problem of extracting the continuum limit values of the
quantities discussed in the previous section.
As  mentioned in the introduction, 
due to the presence of large corrections to 
scaling terms,
this requires a rather 
 non-trivial analysis.  We followed a three step procedure.
\begin{description}
\item{1]}
~~As  first test 
we  tried to fit the data  using only the dominant multiplicative $log$ 
correction keeping into account for $C$ and $\chi$ the possible existence of
bulk constant terms. Hence a one parameter fit for $\xi_2$ and $m$ and a two
parameter fit for $C$ and $\chi$. 
In all  cases we found very high $\chi^2_r$ and, 
even eliminating all the data
except the two couplings nearest to $\beta_c$,
it was
impossible to reach a reasonable confidence level.
This clearly indicates that
additive corrections to the scaling  cannot be neglected.

\item{2]}
~~The second step was to add the first non-universal correction, 
(that of the form 1/log ). 
At the same time we  also included  
the universal corrections evaluated by Salas and Sokal, which are 
of the same order of magnitude and do not add  degrees of freedom 
in the fit. 
The resulting fitting functions are (see sect.~3):
\eq
\chi(t)= a_0 ~+~\Gamma_\pm |t|^{-\frac{7}{6}} (-\log|t| )^{\frac{3}{4}}   
  \left[ 1 + \frac{9}{8}
         \frac{\log(-\log |t|)}{ -\log |t|}
         + \frac{a_1}{ -\log |t|} \right]~ \sk, 
\en
\eq
\xi_2(t)= f_{2,\pm} |t|^{-\frac{2}{3}} (-\log|t| )^{\frac{1}{2}}   
  \left[ 1 + \frac{3}{4}
         \frac{\log(-\log |t|)}{ -\log |t|}
         + \frac{a_1}{ -\log |t|} \right] \sk, 
\en
\eq
m(t) = B |t|^{\frac{1}{12}} (-\log|t| )^{-\frac{1}{8}}   
  \left[ 1 - \frac{3}{16}
         \frac{\log(-\log |t|)}{ -\log |t|}
         + \frac{a_1}{ -\log |t|} \right] \sk. 
\en

\eq
C(t)= a_0 + A_\pm |t|^{-\frac{2}{3}} (-\log|t| )^{-1}   
  \left[ 1 - \frac32
         \frac{\log(-\log |t|)}{ -\log |t|}
         + \frac{a_1}{-\log |t|} \right] \sk,
\en

We performed these
two (or three)
 parameter linear fits, first taking into account  all the data and then
systematically eliminating those farthest from the critical point until an
acceptable reduced $\chi^2$ (namely a $\chi^2$ lower than 1) was reached.
In all  cases, except for the magnetization,
 with this second step we reached 
 an acceptable C.L. and stopped. Notice that in most of the
 cases such acceptable
 C.L. could be reached keeping all the data.
(see the second column of table~8).
We also realized at this stage that  the
 critical amplitudes $A_\pm$  could be obtained in a much more efficient  way by
 looking at the internal energy data. We shall discuss this point in detail in
 the next subsection.

\item{3]}
~~In the next step we added  a next-to-leading non-universal correction. 
Among the various possible terms  we chose  the one  giving,
in the range of values of $\beta$ of our simulations,
  the largest contribution.
We had to resort to this last step only for the magnetization. In this case
there are two competing corrections. The first  one is  the $1/log^2$ term: 
 certainly expected due to the presence of  the marginal field.
But there is also second  possibility:
 thanks to the CFT solution of the
model we know that in the spectrum  there is a subleading 
  magnetic operator to which corresponds a new critical index $\beta'=3/4$.

The two terms have comparable magnitude, but it turns out that
 the subleading
magnetic correction gives a slightly larger contribution in the range of 
interest. So, according to our strategy, we kept only this contribution
 and neglected 
the $1/log^2$ one.  
The resulting scaling function is
\eq
m(t) = B |t|^{\frac{1}{12}} (-\log|t| )^{-\frac{1}{8}}   
  \left[ 1 - \frac{3}{16}
         \frac{\log(-\log |t|)}{ -\log |t|}
         + \frac{a_1}{ -\log |t|} + a_2|t|^{\frac{2}{3}}\right] \sk. 
\en

Adding also the subleading magnetic  term in
the fit  we found an impressive lowering of the $\chi^2$.

\end{description}
The results of these fits are summarized in table~8.
It is important to stress that
all the quoted errors   are  {\sl statistical}.
 Besides them we also 
expect  {\sl systematic} 
errors due to the truncation of the scaling functions to
$O(log|t|)$ or,  in
the case of the magnetization,
to the choice of the next-to-leading non-universal correction. 
In sect. 7.2  we shall discuss  this problem in more detail. 

%
%%%%%%%%%%%%%%%%%%%%%%%%%%%%%%%%%%%%%%%%%%%%%%%%%%%%%%%%%%%%%%%%%%
\begin{table}[htb]
\begin{center}
\begin{tabular}{|c|c|c|c|c|c|c|c|}
\hline
Obs. & $N$ & $\chi^2_r$ & C.L. & Amplitude & $a_0$ & $a_1$  & $a_2$    \\
\hline
$\chi_+$ & 8 & 0.93 & $46 \%$ &$\Gamma_+= 0.0223(14)$&0.05(14)  & 6.5(4) &  \\
$\chi_-$ & 8 & 0.44 & $81 \%$ &$\Gamma_-=  0.00711(10)$&0.02(1) & -1.24(10) &
  \\
$\xi_{2,+}$ & 8 & 0.48 & $82 \%$ &$f_{2,+}=  0.192(4)$ & & 1.35(11) &  \\
$\xi_{2,-}$ & 7 & 0.44 & $82 \%$ & $f_{2,-}=0.088(4) $ && 1.06(27) &  \\
m & 8 & 0.80 & $54 \%$ & $B= 1.1621(11)$  &&  -0.220(6) & -0.144(9)  \\
\hline
\end{tabular}
\caption{\sl Results of the fits for susceptibility, magnetization and
correlation length.
In the second column we report the number of data taken into
account in the fit, in the third column the reduced chi square and 
in the fourth
column the confidence level of the fit. The last four columns contain the best
fit estimates of the parameters of the fit. For the magnetization we also have
the contribution of a next to leading magnetization operator (see text) $a_2$.}
\label{fit1}
\end{center}
\end{table}
%%%%%%%%%%%%%%%%%%%%%%%%%%%%%%%%%%%%%%%%%%%%%%%%%%%%%%%%%%%%%%%%
%

\subsection{The critical amplitudes $A_\pm$}
The most efficient way to obtain the critical amplitudes 
$A_\pm$ is to fit the internal 
energy (for which we have very precise data) instead of the specific heat.
The bulk value of the energy is known (from duality) to be $E(\beta_c)=\frac34$
however, due to the finite size of the lattice that we simulated we must account
for possible small deviations from this asymptotic result. We end up with the
following fitting function
\eq
E(t)= \frac34+a_{-1} + a_0|t| + 3A_\pm |t|^{\frac{1}{3}} (-\log|t| )^{-1}   
  \left[ 1 - \frac{3}{2}
         \frac{\log(-\log |t|)}{-\log |t|}
         + \frac{a_1}{ -\log |t|} \right]~. 
\en
A severe constraint on the results of this fit is represented by duality which
implies $A_+=A_-$ . 
The result of the fits in the low and high T phases are reported in table~9,
where it can be seen that the values of $a_0$, and $A_\pm$ extracted in the two
phases are indeed compatible
within the errors and that $a_{-1}$ is, as expected, very small.
Combining the two estimates of $A_+$ and $A_-$ we extract as our final 
estimate: $A_\pm=1.30(6)$.
%
%%%%%%%%%%%%%%%%%%%%%%%%%%%%%%%%%%%%%%%%%%%%%%%%%%%%%%%%%%%%%
\begin{table}[htb]
\begin{center}
\begin{tabular}{|c|c|c|c|c|c|c|c|}
\hline
Obs. & $N$ & $\chi^2_r$ & C.L. & Amplitude & $a_{-1}$  & $a_0$ & 
$a_1$    \\
\hline
$E_+$ & 7 & 0.67 & $57 \%$ &$A_+=  1.29(5)$  & -0.0001(29) & 
3.0(1.3)&-1.47(48) \\
$E_-$ & 7 & 0.87 & $45 \%$ &$A_-=  1.316(9)$  & -0.0056(5) & 
1.30(18)&-0.93(6) \\
\hline
\end{tabular}
\caption{\sl Results of the fits for the internal energy.}
\label{fit2}
\end{center}
\end{table}
%%%%%%%%%%%%%%%%%%%%%%%%%%%%%%%%%%%%%%%%%%%%%%%%%%%%%%%%%%%%
%
\subsection{Non-universal corrections}
A crucial ingredient to test the reliability of the above fits is given by  
the magnitude of the non-universal corrections. In the range of $\beta$ values
that we studied the log of the reduced temperature $t$ takes 
values which range from $-4$ up to $-5$.
 The non-linear contributions listed in tables~8 and 9 must be 
compared with this reference scale. One easily realizes that for all quantities
these corrections are rather large (they are more or less of the same order
of magnitude of the universal corrections evaluated in ~\cite{ss2})
and in the  particular case of the high
temperature susceptibility  they are very
large.  This suggests that, even if the
fits have a very good confidence level,  caution is needed in assuming the
best fit estimates for the amplitudes which could be affected by  
systematic deviations. 
 Notice that there is no hope to control such  large non-universal
contributions by tuning $\beta$ towards the critical temperature. In fact it
would be necessary to gain at least a factor 10 in log(t) which, as it can be
easily seen, would imply a huge enhancement of $\xi$. 

We tried to estimate the systematic errors which affect our estimates of the
critical amplitude with the following method. We repeated the analysis 
discussed above adding in the fitting functions a term of the form 
$1/(log(t))^2$  (which in the range of values of $t$ that we study 
 is the largest among the correction to scaling terms that we neglect)
 with amplitude equal in modulus to the $a_1$ amplitudes listed in tables 8
 and 9
 and with plus and minus sign. The two resulting values for the critical
 amplitude give a (admittedly rough) idea of the systematic deviations that we
 may expect in our estimates. The results are collected in table 10. By comparing
 with the statistical errors listed in tables 8 and 9. one can see that the
 in all cases the systematic deviations are larger than the statistical errors.

%%%%%%%%%%%%%%%%%%%%%%%%%%%%%%%%%%%%%%%%%%%%%%%%%%%%%%%%%%%%%%%%%%
\begin{table}[htb]
\begin{center}
\begin{tabular}{|c|}
\hline
 Amplitude   \\
\hline
$\Gamma_+= 0.0223(40)$  \\
$\Gamma_-=  0.00711(30)$ \\
$f_{2,+}=  0.192(10)$ \\
$f_{2,-}=0.088(6) $\\
$B= 1.1621(25)$ \\
$A_\pm= 1.30(10)$ \\
\hline
\end{tabular}
\caption{\sl Critical amplitudes with a tentative estimate of the systematic
errors.}
\label{fitsyst}
\end{center}
\end{table}
%%%%%%%%%%%%%%%%%%%%%%%%%%%%%%%%%%%%%%%%%%%%%%%%%%%%%%%%%%%%%%%%

\subsection{Universal Amplitude ratios}

Plugging the values of the critical amplitudes quoted in table~10
in the definitions
(\ref{r1}, \ref{r2}, \ref{r3}, \ref{r4}) 
we find the values for the amplitude ratios reported in the second column of
table~11. The errors quoted in table~11 have been obtained by using the
systematic errors quoted in table~10 and discussed in sect 7.2.

\section{Comparison with field theory predictions}
It is very interesting to compare our results with the 
$R_\chi$, $R_\xi$,$R_c^+$, and $R_\xi^+$
estimates obtained,  
 using the S-matrix 
form-factors approach to the correlation functions, in~\cite{cd}.
It easy to obtain the remaining four ratios by using the following relations:
\eq
R_c^-=\frac{R_c^+}{R_\chi} \sk,\sk
R_\xi^-=\frac{R_\xi^+}{R_\xi} \sk, \sk
R_1=\frac{R_c^-}{(R_\xi^-)^2} \sk,\sk
R_2=\frac{R_c^+}{(R_\xi^+)^2} \sk.
\en
We compare our final estimates and those of \cite{cd} in table~11. We
immediately observe a  good overall agreement. This agreement 
is highly non
trivial since, as discussed above, in our estimates we had to face large
 non universal corrections, while in the predictions of~\cite{cd} only the
 lowest states of the spectrum were taken into account and some small
 discrepancies with the exact results were expected. If we trust in the
  overall agreement that we have found we immediately see that
the major discrepancies between the two sets of data  
are in the two ratios $R_\chi$ and $R_1$, which could both be consequences of a
biased estimate of $\Gamma_-$. 
It would be important to understand
the reason of this discrepancy.
In this respect it is worthwhile to notice 
that $\chi_l$ is the
only observable for which the non-universal correction has the opposite sign
with respect to the universal additive one. 
%
%%%%%%%%%%%%%%%%%%%%%%%%%%%%%%%%%%%%%%%%%%%%%%%%%%%%%
\begin{table}[htb]
\begin{center}
\begin{tabular}{|c|c|c|}
\hline
Ratio & this work &  ref.~\cite{cd} \\
\hline
$ R_\chi   $&$  3.14(70)    $ & 4.013  \\
$ R_\xi    $&$  2.19(26)    $ & 1.935  \\
$ R_1      $&$  0.68(13)   $ &0.4539  \\
$ R_2      $&$  0.44(13)   $ & 0.4845 \\
$ R_c^+    $&$  0.021(5)  $ & 0.0204 \\
$ R_c^-    $&$  0.0068(9) $ &0.0051  \\
$ R_\xi^+  $&$  0.220(20)   $ & 0.2052 \\
$ R_\xi^-  $&$  0.100(10)    $ & 0.1060 \\
\hline
\end{tabular}
\caption{\sl Comparison between our estimates for the universal 
amplitude ratios and those of ref.~{\protect \cite{cd}}. }
\label{fit8}
\end{center}
\end{table}
%%%%%%%%%%%%%%%%%%%%%%%%%%%%%%%%%%%%%%%%%%%%%%%%%%
%
\section{Concluding remarks}

The aim of this paper was 
 to test the recent predictions of~\cite{cd}
for various amplitude ratios in the 4 state Potts model,
with the results of a high precision Monte Carlo simulation.
 
We made four tests on the results of our simulations: 
\begin{itemize}
\item
Comparison with exact results for small lattices.
\item
Comparison with the results of ~\cite{ss1,ss2} at the critical point.
\item
Comparison with low temperature series.
\item
Agreement with the duality relations.
\end{itemize}
All these tests were successfully passed.

In looking at 
the scaling behaviour of our
observables, we had to face a major problem,
 due to the presence of a marginal field in the spectrum. 
In performing the analysis we used the recent results of Salas and 
Sokal on the
universal additive $log~log/log$ correction terms, and found in our fits 
the same behaviour and the same features that they reported in~\cite{ss2}
where they looked at the finite size corrections at the critical point.

We found a relatively good overall agreement with the predictions of
Cardy and Delfino
with the exception of two ratios involving the low temperature susceptibility. 
It remains an open problem to find  a more efficient 
way of  dealing, in the analysis of Monte Carlo  data,
 with corrections originated by the presence of
 marginal operators. These contributions  are the probable  main
cause of these  discrepancies.

\vskip 0.5cm
{\bf  Acknowledgements}

We thank   J.L.Cardy, G.Delfino, P.Dorey,  F.Gliozzi and B.Nienhuis  
for helpful 
discussions. 
This work was partially supported by the 
European Commission TMR programme ERBFMRX-CT96-0045.
\vskip 1cm

\end{document}